

\newdimen\fullhsize
\newdimen\hstitle
\hstitle=\hsize 
\newdimen\hsbody
\hsbody=\hsize 
\newdimen\hbodyoffset
\hbodyoffset=\hoffset 
\newbox\leftpage

\magnification=1000  
\baselineskip=14pt
\global\hstitle=9truein\global\hsbody=4.75truein
\global\vsize=7truein\global\voffset=-.31truein
\global\hoffset=-0.54in\global\hbodyoffset=-.54truein
\global\fullhsize=10truein

\global\newcount\secno \global\secno=0
\global\newcount\appno \global\appno=0
\global\newcount\meqno \global\meqno=1
\global\newcount\chapno \global\chapno=0
\global\newcount\figno \global\figno=0
\def\secta{\global\advance\secno by1
\meqno=1}

\def\chapter#1{\global\advance\chapno by1
\global\secno=0
\vfill\eject
\centerline{\titlesize\bf Chapter~\the\chapno. }
\bigskip\centerline{\titlesize\bf #1.}
\bigskip}

\def\Section#1{\global\advance\secno by1\relax\global\meqno=1
\bigbreak\bigskip
{\noindent\sectsize \bf \the\chapno.\the\secno ~#1.}%
\par\nobreak\medskip\nobreak}
\def\tagsection#1{\edef#1{\the\chapno.\the\secno}%
\ifWritingAuxFile\immediate\write\auxfile{\noexpand\xdef\noexpand#1{#1}}\fi%
}
\def\section{\Section}

\def\romappno{\uppercase\expandafter{\romannumeral\appno}}
\def\Appendix#1{\global\advance\appno by1\relax\global\meqno=1\global\secno=0
\bigbreak\bigskip
\centerline{\twelvepoint \bf Appendix %
A. #1}%
\par\nobreak\medskip\nobreak}
\def\tagappendix#1{\edef#1{\romappno}%
\ifWritingAuxFile\immediate\write\auxfile{\noexpand\xdef\noexpand#1{#1}}\fi%
}
\def\appendix{\Appendix}

\def\eqn#1{
\ifnum\secno>0

\eqno(\the\chapno.\the\secno.\the\meqno)\xdef#1{\the\chapno.\the\secno.\the\meqno}%
     \global\advance\meqno by1
\else\ifnum\appno>0
  \eqno({\rm\romappno}.\the\meqno)\xdef#1{\romappno.\the\meqno}%
     \global\advance\meqno by1
\else
  \eqno(\the\meqno)\xdef#1{\the\meqno}%
     \glovbal\advance\meqno by1
\fi\fi%
\ifWritingAuxFile\immediate\write\auxfile{\noexpand\xdef\noexpand#1{#1}}\fi%
}
\def\equn{
\ifnum\secno>0
  \eqno(\the\chapno.\the\secno.\the\meqno)%
     \global\advance\meqno by1
\else\ifnum\appno>0
  \eqno(A.\the\meqno)%
     \global\advance\meqno by1
\else
  \eqno(\the\meqno)%
     \global\advance\meqno by1
\fi\fi%
}
\def\defeqn#1{
\ifnum\secno>0
  \xdef#1{\the\chapno.\the\secno.\the\meqno}%
     \global\advance\meqno by1
\else\ifnum\appno>0
  \xdef#1{\romappno.\the\meqno}%
     \global\advance\meqno by1
\else
  \xdef#1{\the\meqno}%
     \global\advance\meqno by1
\fi\fi%
\ifWritingAuxFile\immediate\write\auxfile{\noexpand\xdef\noexpand#1{#1}}\fi%
}
\def\anoneqn{
\ifnum\secno>0
  \eqno(\the\chapno.\the\secno.\the\meqno)%
     \global\advance\meqno by1
\else\ifnum\appno>0
  \eqno({\rm\romappno}.\the\meqno)%
     \global\advance\meqno by1
\else
  \eqno(\the\meqno)%
     \global\advance\meqno by1
\fi\fi%
}
\def\mfig#1#2{\global\advance\figno by1%
\relax#1\the\figno\edef#2{\the\figno}%
\ifWritingAuxFile\immediate\write\auxfile{\noexpand\xdef\noexpand#2{#2}}\fi%
}

\catcode`@=11 

\font\ninerm=cmr9
\font\eightrm=cmr8
\font\sixrm=cmr6

\def\loadtrueseventeenpoint{
 \font\seventeenrm=cmr10 at 17.28truept
 \font\seventeeni=cmmi10 at 17.28truept
 \font\seventeenbf=cmbx10 at 17.28truept
 \font\seventeenit=cmti10 at 17.28truept
 \font\seventeensl=cmsl10 at 17.28truept
 \font\seventeensy=cmsy10 at 17.28truept
}
\def\loadfourteenpoint{
\font\fourteenrm=cmr10 at 14.4pt
\font\fourteeni=cmmi10 at 14.4pt
\font\fourteenit=cmti10 at 14.4pt
\font\fourteensl=cmsl10 at 14.4pt
\font\fourteensy=cmsy10 at 14.4pt
\font\fourteenbf=cmbx10 at 14.4pt
}
\def\loadtruetwelvepoint{
\font\twelverm=cmr10 at 12truept
\font\twelvei=cmmi10 at 12truept
\font\twelveit=cmti10 at 12truept
\font\twelvesl=cmsl10 at 12truept
\font\twelvesy=cmsy10 at 12truept
\font\twelvebf=cmbx10 at 12truept
}

\font\ninei=cmmi9
\font\eighti=cmmi8
\font\sixi=cmmi6
\skewchar\ninei='177 \skewchar\eighti='177 \skewchar\sixi='177

\font\ninesy=cmsy9
\font\eightsy=cmsy8
\font\sixsy=cmsy6
\skewchar\ninesy='60 \skewchar\eightsy='60 \skewchar\sixsy='60

\font\ninebf=cmbx9
\font\eightbf=cmbx8
\font\sixbf=cmbx6

\font\ninett=cmtt9
\font\eighttt=cmtt8

\hyphenchar\tentt=-1 
\hyphenchar\ninett=-1
\hyphenchar\eighttt=-1

\font\ninesl=cmsl9
\font\eightsl=cmsl8

\font\nineit=cmti9
\font\eightit=cmti8


\newskip\ttglue
\def\tenpoint{\def\rm{\fam0\tenrm}%
  \textfont0=\tenrm \scriptfont0=\sevenrm \scriptscriptfont0=\fiverm
  \textfont1=\teni \scriptfont1=\seveni \scriptscriptfont1=\fivei
  \textfont2=\tensy \scriptfont2=\sevensy \scriptscriptfont2=\fivesy
  \textfont3=\tenex \scriptfont3=\tenex \scriptscriptfont3=\tenex
  \def\it{\fam\itfam\tenit}\textfont\itfam=\tenit
  \def\sl{\fam\slfam\tensl}\textfont\slfam=\tensl
  \def\bf{\fam\bffam\tenbf}\textfont\bffam=\tenbf \scriptfont\bffam=\sevenbf
  \scriptscriptfont\bffam=\fivebf
  \normalbaselineskip=12pt
  \let\sc=\eightrm
  \let\big=\tenbig
  \setbox\strutbox=\hbox{\vrule height8.5pt depth3.5pt width\z@}%
  \normalbaselines\rm}

\def\twelvepoint{\def\rm{\fam0\twelverm}%
  \textfont0=\twelverm \scriptfont0=\ninerm \scriptscriptfont0=\sevenrm
  \textfont1=\twelvei \scriptfont1=\ninei \scriptscriptfont1=\seveni
  \textfont2=\twelvesy \scriptfont2=\ninesy \scriptscriptfont2=\sevensy
  \textfont3=\tenex \scriptfont3=\tenex \scriptscriptfont3=\tenex
  \def\it{\fam\itfam\twelveit}\textfont\itfam=\twelveit
  \def\sl{\fam\slfam\twelvesl}\textfont\slfam=\twelvesl
  \def\bf{\fam\bffam\twelvebf}\textfont\bffam=\twelvebf
                                           \scriptfont\bffam=\ninebf
  \scriptscriptfont\bffam=\sevenbf
  \normalbaselineskip=12pt
  \let\sc=\eightrm
  \let\big=\tenbig
  \setbox\strutbox=\hbox{\vrule height8.5pt depth3.5pt width\z@}%
  \normalbaselines\rm}

\def\fourteenpoint{\def\rm{\fam0\fourteenrm}%
  \textfont0=\fourteenrm \scriptfont0=\tenrm \scriptscriptfont0=\sevenrm
  \textfont1=\fourteeni \scriptfont1=\teni \scriptscriptfont1=\seveni
  \textfont2=\fourteensy \scriptfont2=\tensy \scriptscriptfont2=\sevensy
  \textfont3=\tenex \scriptfont3=\tenex \scriptscriptfont3=\tenex
  \def\it{\fam\itfam\fourteenit}\textfont\itfam=\fourteenit
  \def\sl{\fam\slfam\fourteensl}\textfont\slfam=\fourteensl
  \def\bf{\fam\bffam\fourteenbf}\textfont\bffam=\fourteenbf%
  \scriptfont\bffam=\tenbf
  \scriptscriptfont\bffam=\sevenbf
  \normalbaselineskip=17pt
  \let\sc=\elevenrm
  \let\big=\tenbig
  \setbox\strutbox=\hbox{\vrule height8.5pt depth3.5pt width\z@}%
  \normalbaselines\rm}

\def\seventeenpoint{\def\rm{\fam0\seventeenrm}%
  \textfont0=\seventeenrm \scriptfont0=\fourteenrm \scriptscriptfont0=\tenrm
  \textfont1=\seventeeni \scriptfont1=\fourteeni \scriptscriptfont1=\teni
  \textfont2=\seventeensy \scriptfont2=\fourteensy \scriptscriptfont2=\tensy
  \textfont3=\tenex \scriptfont3=\tenex \scriptscriptfont3=\tenex
  \def\it{\fam\itfam\seventeenit}\textfont\itfam=\seventeenit
  \def\sl{\fam\slfam\seventeensl}\textfont\slfam=\seventeensl
  \def\bf{\fam\bffam\seventeenbf}\textfont\bffam=\seventeenbf%
  \scriptfont\bffam=\fourteenbf
  \scriptscriptfont\bffam=\twelvebf
  \normalbaselineskip=21pt
  \let\sc=\fourteenrm
  \let\big=\tenbig
  \setbox\strutbox=\hbox{\vrule height 12pt depth 6pt width\z@}%
  \normalbaselines\rm}

\def\ninepoint{\def\rm{\fam0\ninerm}%
  \textfont0=\ninerm \scriptfont0=\sixrm \scriptscriptfont0=\fiverm
  \textfont1=\ninei \scriptfont1=\sixi \scriptscriptfont1=\fivei
  \textfont2=\ninesy \scriptfont2=\sixsy \scriptscriptfont2=\fivesy
  \textfont3=\tenex \scriptfont3=\tenex \scriptscriptfont3=\tenex
  \def\it{\fam\itfam\nineit}\textfont\itfam=\nineit
  \def\sl{\fam\slfam\ninesl}\textfont\slfam=\ninesl
  \def\bf{\fam\bffam\ninebf}\textfont\bffam=\ninebf \scriptfont\bffam=\sixbf
  \scriptscriptfont\bffam=\fivebf
  \normalbaselineskip=11pt
  \let\sc=\sevenrm
  \let\big=\ninebig
  \setbox\strutbox=\hbox{\vrule height8pt depth3pt width\z@}%
  \normalbaselines\rm}

\def\eightpoint{\def\rm{\fam0\eightrm}%
  \textfont0=\eightrm \scriptfont0=\sixrm \scriptscriptfont0=\fiverm%
  \textfont1=\eighti \scriptfont1=\sixi \scriptscriptfont1=\fivei%
  \textfont2=\eightsy \scriptfont2=\sixsy \scriptscriptfont2=\fivesy%
  \textfont3=\tenex \scriptfont3=\tenex \scriptscriptfont3=\tenex%
  \def\it{\fam\itfam\eightit}\textfont\itfam=\eightit%
  \def\sl{\fam\slfam\eightsl}\textfont\slfam=\eightsl%
  \def\bf{\fam\bffam\eightbf}\textfont\bffam=\eightbf \scriptfont\bffam=\sixbf%
  \scriptscriptfont\bffam=\fivebf%
  \normalbaselineskip=9pt%
  \let\sc=\sixrm%
  \let\big=\eightbig%
  \setbox\strutbox=\hbox{\vrule height7pt depth2pt width\z@}%
  \normalbaselines\rm}

\def\tenbig#1{{\hbox{$\left#1\vbox to8.5pt{}\right.\n@space$}}}
\def\ninebig#1{{\hbox{$\textfont0=\tenrm\textfont2=\tensy
  \left#1\vbox to7.25pt{}\right.\n@space$}}}
\def\eightbig#1{{\hbox{$\textfont0=\ninerm\textfont2=\ninesy
  \left#1\vbox to6.5pt{}\right.\n@space$}}}

\def\footnote#1{\edef\@sf{\spacefactor\the\spacefactor}#1\@sf
      \insert\footins\bgroup\eightpoint
      \interlinepenalty100 \let\par=\endgraf
        \leftskip=\z@skip \rightskip=\z@skip
        \splittopskip=10pt plus 1pt minus 1pt \floatingpenalty=20000
        \smallskip\item{#1}\bgroup\strut\aftergroup\@foot\let\next}
\skip\footins=12pt plus 2pt minus 4pt 
\dimen\footins=30pc 

\newinsert\margin
\dimen\margin=\maxdimen

\loadtruetwelvepoint 
\loadtrueseventeenpoint
\catcode`\"=\active

\def\eatOne#1{}
\def\ifundef#1{\expandafter\ifx%
\csname\expandafter\eatOne\string#1\endcsname\relax}


\global\newcount\refno \global\refno=1
\newwrite\rfile
\newlinechar=`\^^J
\def\ref#1#2{\the\refno\nref#1{#2}}
\def\nref#1#2{\xdef#1{\the\refno}%
\ifnum\refno=1\immediate\openout\rfile=refs.tmp\fi%
\immediate\write\rfile{\noexpand\item{[\noexpand#1]\ }#2.}%
\global\advance\refno by1}
\def\lref#1#2{\the\refno\xdef#1{\the\refno}%
\ifnum\refno=1\immediate\openout\rfile=refs.tmp\fi%
\immediate\write\rfile{\noexpand\item{[\noexpand#1]\ }#2\semi}%
\global\advance\refno by1}
\def\cref#1{\immediate\write\rfile{#1\semi}}

\def\semi{;\hfil\noexpand\break}

\newif\ifWritingAuxFile
\newwrite\auxfile
\def\SetUpAuxFile{%
\xdef\auxfileName{\jobname.aux}%
\openin1 \auxfileName \ifeof1\message{No file \auxfileName; I'll create one.
}\else\closein1\relax\input\auxfileName\fi%
\WritingAuxFiletrue%
\immediate\openout\auxfile=\auxfileName}

\def\L{\left(}

\def\ref#1#2{\nref#1{#2}}
\overfullrule 0pt
\hfuzz 52pt
\hsize 6.50 truein
\vsize 9.5 truein

\loadfourteenpoint
\newcount\meqno
\newcount\secno
\meqno=0
\secno=0

\def\put#1{\global\edef#1{(\the\chapno.\the\secno.\the\meqno)}     }

\twelvepoint
\baselineskip 24pt
\lineskip 24pt
\hsize 15.0truecm
\hoffset 1.0truecm

\def\k{\kappa}
\def\M{$\overline{\hbox{MS}}$ }

\def\b{\beta}
\def\p{\phi}
\def\g{\gamma}
\def\d{\partial}
\def\pd{\d}
\def\m{\mu}
\def\l{\lambda}
\def\h{\hbox{${1\over2}$}}
\def\D{{\cal D}}
\def\L{\Lambda}

\def\f{(4\pi)^2}
\hfill DIAS STP 94-09\par
\hfill hep-th 9404085\par
\hfill April 1994\par
\hfill revised September 1994\par
\line{\hfill}
\centerline{\seventeenpoint\bf Multi-scale Renormalisation Group
Improvement}
\line{\hfill}
\centerline{\seventeenpoint\bf of the Effective Potential}
\line{\hfill}

\line{\hfill}
\line{\hfill}
\centerline{ Christopher Ford}
\line{\hfill}
\centerline{ Dublin Institute for Advanced Studies, }
\centerline{10, Burlington Road, Dublin 4, Ireland.}
\line{\hfill}
\line{\hfill}
\centerline{\fourteenpoint\bf Abstract}
\line{\hfill}
Using the renormalisation group and a conjecture concerning the
perturbation series for the effective potential, the leading
logarithms in the effective potential are exactly summed for $O(N)$
scalar and Yukawa theories.

\line{\hfill}
\line{\hfill}
\vfill\eject
\centerline{\fourteenpoint\bf 1. Introduction}
\line{\hfill}

The effective potential (EP) is a very useful tool for the study of
spontaneous symmetry breaking in field theory. However in many applications
the perturbative loop expansion is inadequate even if the couplings are
``small''. This is due to the presence of logarithmic terms, like
$\ln(\p/\m)$, in the
perturbation series which restrict the range of $\p$ values (here $\p$
is some generic scalar field, and $\m$ is the renormalisation scale)
 where perturbation theory is credible.
In applications where one needs to survey the EP for a wide  range of
$\p$ values  (eg.
vacuum stability analyses in the Standard Model [1,2,3]), these logarithms must
be
dealt with.  Of course, one can simply let the parameters run,
 and calculate the EP at $\m=\p$. Provided
these running couplings remain perturbative at this scale, one can
drastically extend the scope of perturbation theory. An alternative
way of thinking about renormalisation group (RG) improvement, is
to view it as a reorganization of the perturbation series, in which
the first term is the sum of all the \sl
leading logarithms, \rm the second term represents the
sub-leading logarithms, and so on. The leading logarithms  are
terms of the form $\hbar^n\ln(M_1(\p)/\m)\ln(M_2(\p)/\m)...
\ln(M_n(\p)/\m)$, and represent the most ``dangerous''
logarithmic terms at each order in perturbation theory (note that
the tree potential is counted as a leading logarithm). The
sub-leading logarithms are proportional to \footnote{$\dagger$}
{We retain the factors of $\hbar$ in all equations so that the
reader can easily distinguish leading from sub-leading
contributions.}
$\hbar^{n+1}
\ln(M_1/\m)...\ln(M_n/\m)$.
In  dimensional regularisation, the most divergent $n$-loop terms
one encounters (in a model with a single mass scale $m$) are
proportional to $\hbar^n(m/\m)^\epsilon/\epsilon^n$
 . When these terms are expanded in powers of $\epsilon$ finite
terms proportional to $\hbar^n\ln^n(m/\m)$ are generated (in a
theory with
two scales $m_1$ and $m_2$ terms of the form $\hbar^n\ln^p(m_1/\m)
\ln^{n-p}(m_2/\m)$ will appear). Therefore
it is only the most divergent pieces of the Feynman diagrams that
contribute to the leading logarithms.

Renormalisation group improved
potentials were first considered in the context of massless models
by Coleman and Weinberg [4]. Recently [2,5,6] it has been demonstrated that
this treatment also works in the massive case provided one takes
into account the running of the vacuum energy (or cosmological
constant).
 However, when there is more than one mass scale present
it is less clear how to proceed; no choice of $\m$ will kill all the
logarithms.  This is not usually important provided that the
logarithms of the scale ratio's are \lq\lq small''. However, in certain
cases of interest these logarithms are large (for example
$\ln(M_{\hbox{GUT}}/M_{\hbox{electroweak}})\approx 30$). Even in
situations where these logarithms are not so large the scope of
perturbation theory is still reduced, for example consider a two
scale model with $m_1>m_2$, then perturbative credibility requires
that the $\l_i\ln(m_1/m_2)$ be \lq\lq small'' in \sl addition \rm to the usual
requirement that just the couplings $\l_i$ are small. If we could
fully
sum the multi-scale leading logarithms we should have an
approximation that is useful despite the existence of widely
differing scales.
 In ref. [7] it was argued that the decoupling theorem
could be used to obtain approximations to the multi-scale leading
logarithms within the \M  scheme. Alternatively, in ref. [8] it
was found that some of the problems associated with RG improvement
are absent in a modified mass-dependent scheme, although the RG
equation is difficult to work with in such schemes.
 In this paper it is suggested that it
may be possible to \sl exactly \rm sum the leading logarithms in a
general theory, using a mass-independent renormalisation scheme.
 Explicit formulae are presented in the case of
$O(N)$ scalar and Yukawa models. The results presented here depend
on an unproven conjecture, however the conjecture has been checked
to two loops.

The outline of this paper is as follows. In section 2, we review the
leading logarithms calculation for ordinary massive $\p^4$ theory
(the result is later used as a boundary condition for the full
$O(N)$ calculations). In section 3 we present our method and
calculation of the leading logarithms in the $O(N)$ scalar $\p^4$
theory. We apply our method to the more complicated Yukawa model in
section 4, and in section 5 we conclude with a discussion of the
general validity of the method.

\line{\hfill}
\line{\hfill}

\centerline{\fourteenpoint\bf 2. Massive $\p^4$ Theory}
\line{\hfill}
Consider massive $\p^4$ theory in four dimensions defined by the
Lagrangian
$${\cal
L}={1\over2}(\d_\m\p)^2-{1\over2}m^2\p^2-{\l\over{24}}\p^4-\L.\eqno(2.1)$$
Here $\L$ is a \lq\lq cosmological constant'' term. Assuming the EP,
 $V(\p)$, is independent of the renormalisation
scale, $\m$, for fixed values of the bare parameters, one obtains
the following RG equation
$$ \D V=0,\eqno (2.2)$$
where
$$\D=\m{\d\over{\d \m}}+\b_\l{\d\over{\d\l}}+\b_{m^2}{\d\over{\d m^2}}-
\g \p{\d\over{\d\p}}+\b_\L{\d\over{\d\L}}.\eqno(2.3)$$
Here $\b_\l$, $\b_{m^2}$ and $\b_{\L}$ are the coupling constant,
 mass squared and
cosmological constant beta
functions, respectively, and $\g$ is the anomalous dimension.
The tree potential can be read off from the Lagrangian,
$$V^{(0)}={1\over2}m^2\p^2+{\l\over{24}}\p^4+\L,\eqno(2.4)$$
and the one-loop potential is easily calculated [4,9,10], the result in \M
reads
$$V^{(1)}={\hbar(m^2+\h\l\p^2)^2\over{4(4\pi)^2}}\left(
\ln{m^2+\h\l\p^2\over{\m^2}}-{3\over2}\right).\eqno(2.5)$$
The one-loop RG functions are
$$\b^{(1)}_\l={3\hbar\l^2\over{(4\pi)^2}},\qquad
\b^{(1)}_{m^2}={\hbar m^2\l\over{(4\pi)^2}},\qquad
\b^{(1)}_{\L}={\hbar m^4\over{2\f}},\qquad\g^{(1)}=0.\eqno(2.6)$$

Applying the method of characteristics to eq. (2.2)
$$V(\l,m^2,\m,\L,\p)= V(\bar \l,\bar m^2,\bar \m,\bar\L,\bar
\p),\eqno(2.7)$$
where the ``running'' parameters satisfy
 $$\eqalign{
\hbar{d\bar\m\over{dt}}=&\bar\m,\cr
\hbar{d\bar\L\over{dt}}=&\b_\L(\bar\l,\bar m^2),}
\quad
\eqalign{\hbar{d\bar\l\over{dt}}=&\b_\l(\bar \l),\cr
\hbar{d\bar\p\over{dt}}=&-\g(\bar\l)\bar\p,\cr}
\quad
\eqalign{
\hbar{d\bar m^2\over{dt}}=&\b_{m^2}(\bar m^2,\bar \l),\cr
&\cr &\cr }\eqno(2.8)$$
 and $\bar \m(t=0)=\m$,
$\bar\l(t=0)= \l$, $\bar{m^2}(t=0)=m^2$, $\bar\L(t)=\L$,
 $\bar \p(t=0)=\p$.
The idea of RG improvement is that via a judicious choice of $t$,
 one  can evaluate the right hand side of eq. (2.7) perturbatively
even if large logarithms render the left hand side non-perturbative.
This method yields approximations to the EP which are useful for a
much wider range of $\p$ values than the conventional loop
expansion.
 The obvious choice would be to choose $t$ so as to remove \sl all
\rm the logarithms on the right hand side of eq. (2.7) (note that this
is only possible because there is only one kind of logarithm,
namely $\ln[(m^2+\h\l\p^2)/\m^2]$,  in the perturbation
series). That is, $t$ is chosen so that
$${\bar{m^2}(t)+\h\bar\l(t){\bar \p(t)}^2\over{{\bar\m(t)}^2}}=1.
\eqno(2.9)$$
While eq. (2.9) seems the most natural choice, it is awkward to work
with (even in the one-loop approximation).
A rather less implicit choice is given by
$$t={\hbar\over2}\ln{m^2+\h\l\p^2\over{\m^2}}.\eqno(2.10)$$
Note that this choice does \sl not \rm kill the logarithms on the
right hand side of eq. (2.7), however it allows one to explicitly sum
the leading (and subleading, etc.) logarithms in the EP [2,5,6].
 With this choice of $t$, $\bar\m^2(t)=\m^2\exp(2t/\hbar)=m^2+\h\l\p^2$
which is independent of
$\m$; all the $\m$ dependence is carried by $\bar\l(t)$, $\bar
m^2(t)$, $\bar\L(t)$ and
$\bar\p(t)$.
Solving eqs. (2.8) using the one-loop RG functions,
$$\eqalign{
 \bar\l(t)&=\l\left(1-{3\l t\over{(4\pi)^2}}\right)^{-1}+{\cal O}
(\hbar),\cr
\bar{m^2}(t)&=m^2\left(1-{3\l t\over{(4\pi)^2}}\right)^{-1/3}+
{\cal O}(\hbar),\quad\bar\p(t)=\p+{\cal O}(\hbar),\cr
\bar\L(t)&= \L-{m^4\over{2\l}}\left[
\left(1-{3\l t\over\f}\right)^{1/3}-1\right]+{\cal O}(\hbar).
\cr}\eqno(2.11)$$
Inserting these into the right hand side of eq. (2.7), one finds
$$\eqalign{V=&{\l\over{24}}\p^4\left(1-{3\l t\over{(4\pi)^2}}\right)^{-1}
+{1\over2}m^2\p^2\left(1-{3\l t\over{(4\pi)^2}}\right)^{-1/3}
-{m^4\over{2\l}}\left(1-{3\l t\over{(4\pi)^2}}\right)^{1/3}\cr
&+{m^4\over{2\l}}+\L+{\cal O}(\hbar).\cr}\eqno (2.12)$$
With the choice of $t$ given by (2.10), eq. (2.12) gives the sum of the
leading logarithms in the EP (which was first obtained
by Kastening [5]). The ${\cal O}(\hbar)$ term represents
the sub-leading, sub-sub-leading, etc. contributions to the EP. In
general, to perform the leading (sub-leading, ...) logarithms
expansion, one must expand the right hand side of eq. (2.7) in powers
of $\hbar$ \sl but retaining all orders \rm in $t$. However, in this
paper we just concentrate on the leading logarithms summation, which
(in the single mass scale case) just amounts to substituting the
one-loop running parameters into the tree level potential.
\line{\hfill}
\line{\hfill}

\centerline{\fourteenpoint \bf 3. $O(N)$ symmetric $\p^4$ theory}
\line{\hfill}
Consider the theory defined by the Lagrangian
$${\cal
L}={1\over2}(\d_\m\p)^2-{1\over2}m^2\p^2-{\l\over{24}}\p^4-\L,\eqno
(3.1)$$
where $\p^2=\p_i\p_i \qquad (i=1,..,N)$, and $\p_i$ is an
$N$-component scalar field. Although this model has $N$ scalar
fields we can exploit the $O(N)$ invariance to write the EP,
$V(\p_i)$, as a function of $\p$ only. The tree-level potential is
simply
$$V^{(0)}={1\over2}m^2\p^2+{\l\over{24}}\p^4+\L,\eqno (3.2)$$
note that this is independent of $N$, which will be exploited when
summing the leading logarithms. The one-loop potential reads
$$(4\pi)^2V^{(1)}={\hbar\over4}H^2\left(\ln{H\over{\m^2}}-{3\over2}\right)
+{\hbar\over4}(N-1)G^2\left(\ln{G\over{\m^2}}-{3\over2}\right),\eqno (3.3)$$
where
$$ H=m^2+\h\l\p^2,\qquad G=m^2+\hbox{${1\over6}$}\l\p^2.\eqno (3.4)$$
The two-loop potential is also known for this model [11]. Note that
for $N\neq 1$ we have \sl two \rm distinct logarithms in the
perturbation series, so unlike the $N=1$ case no choice of $\m$ will
remove all the logarithms. Of course we can write the second
logarithm in terms of the first
$$\ln {G\over{\m^2}}=\ln{H\over{\m^2}}+\ln{G\over H},\eqno (3.5)$$
then provided $\ln(H/G)$ is ``small'' we can sum up the $\ln(H/\m^2)$
terms in the same fashion as the $N=1$ case [12]. Here we will
consider summation of \sl both \rm logarithms. On physical grounds,
one could argue that this is not really necessary, since although
the EP contains two logarithms, there is really only one \sl physical
\rm scale in the theory. For if $m^2>0$, we have $N$ particles of the
same mass, and if  $m^2<0$ we have one massive particle and $N-1$
\sl massless \rm Goldstone bosons. However, if we can sum up the two
logarithms in this simple model we should be able to apply the method to
cases where there is more than one physical scale (such as the Yukawa model
treated in the next section).

Consider the sum of the leading logarithms
$$\eqalign{
L=&{1\over2}m^2\p^2+{\l\over{24}}\p^4+\L+{\hbar H^2\over{4(4\pi)^2}}
\ln{H\over{\m^2}}
+{\hbar(N-1)G^2\over{4(4\pi)^2}}\ln{G\over{\m^2}}\cr
&+{\hbar^2(\l\p)^2H\over{8(4\pi)^4}}\ln^2{H\over{\m^2}}
+{\hbar^2\l(N^2-1)G^2\over{24(4\pi)^4}}\ln^2{G\over{\m^2}}\cr
&+{\hbar^2(N-1)(\l\p)^2\over{72(4\pi)^4}}\left(
(2G-H)\ln^2{G\over {\m^2}}+2H\ln{H\over{\m^2}}\ln{G\over{\m^2}}\right)\cr
&+{\hbar^2\l H^2\over8{(4\pi)^4}}\ln^2{H\over{\m^2}}+{\hbar^2\l
(N-1)HG\over{12(4\pi)^4}}
\ln{H\over{\m^2}}\ln{G\over{\m^2}}+{\cal O}(\hbar^3),\cr
}\eqno(3.6)
$$
(the two-loop terms were obtained in [11]).
Note that the $-3/2$ terms in $V^{(1)}$  are not included, since
they are counted as sub-leading logarithms which we do not attempt to
sum here. Using the RG alone it is impossible to compute $L$
exactly (this is because the operator $\m\d/\d\m$ cannot distinguish
between $\ln(H/\m^2)$ and $\ln(G/\m^2)$). Now the crucial point is to
notice that if we set $\m^2=G$ in eq. (3.6), $L$ reduces to the $N=1$ case
(at least to the two-loop level).
Note that this is \sl not \rm true for the sub-leading logarithms,
but seems to hold for the leading logarithms. We know (exactly) what
$L$ is for $N=1$ (eq. (2.12)), thus we assume
$$\eqalign{
L(\m^2=G)=&{\l\over{24}}\p^4\left(
1-{3\l\hbar\over{2(4\pi)^2}}\ln{H\over G}\right)^{-1}
+{1\over2}m^2\p^2\left(
1-{3\l\hbar\over{2(4\pi)^2}}\ln{H\over G}\right)^{-1/3}\cr
&-{m^4\over{2\l}}\left(
1-{3\l\hbar\over{2(4\pi)^2}}\ln{H\over G}\right)^{1/3}
+{m^4\over{2\l}}+\L.\cr}\eqno(3.7)$$
So if we solve the RG for the $O(N)$ EP using the $N=1$ formula as a
boundary condition at $\m^2=G$ we should be able to compute $L$
exactly.

The one-loop RG functions for this model are
$$\b_\l^{(1)}={N+8\over{3\f}}\hbar\l^2,\qquad\b^{(1)}_{m^2}={N+2\over{3\f}}
\hbar\l
m^2,
\qquad \b^{(1)}_{\L}={N\hbar m^4\over{2\f}},\qquad\g^{(1)}=0,
\eqno(3.8)$$
which gives the following one-loop running couplings
\def\br{\left(1-{(N+8)\l t\over{3(4\pi)^2}}\right)}
\def\bbr{\left(1-{3\l s\over{(4\pi)^2}}-{(N-1)\l t\over{3(4\pi)^2}}
\right)}
\def\lbbr{\left(1-\hbox{${3\l s\over{(4\pi)^2}}$}-\hbox{${(N-1)\l
t\over{3(4\pi)^2}}$}\right)}
$$\eqalign{\bar\l(t)&=\l\br^{-1},\cr
 \bar{m^2}(t)&=m^2\br^{-{N+2\over{N+8}}},\qquad\bar\p(t)=\p,\cr
\bar\L(t)&=\L+{3Nm^4\over{2(N-4)\l}}\left[
\br^{-{N-4\over{N+8}}}-1\right].\cr}\eqno(3.9)$$
Now choose
$$t={\hbar\over2}\ln{G\over{\m^2}},\eqno(3.10)$$
which means that $t=0$ corresponds to $G=\m^2$, and insert eqs.
(3.9) into eq. (2.7).
The RG ``improved'' potential obtained using the one-loop running
parameters together with the boundary condition eq. (3.7) is
$$\eqalign{V_{\hbox{improved}}=&{\bar\l\over{24}}\bar\p^4
\left(1-{3\bar\l\hbar\over{2\f}}\ln{\bar H\over{\bar G}}
\right)^{-1}+{1\over2}\bar m^2\bar\p^2\left(
1-{3\bar\l\hbar\over{2\f}}\ln{\bar H\over{\bar G}}\right)^{-1/3}\cr
&-{\bar m^4\over{2\bar\l}}\left(
1-{3\bar\l\hbar\over{2\f}}\ln{\bar H\over{\bar
G}}\right)^{1/3}+{\bar m^4\over{2\bar \l}}+\bar\L,\cr}\eqno(3.11)$$
where $\bar H=\bar m^2+\h\bar \l \bar\p^2$ and
$\bar G=\bar m^2+\hbox{${1\over6}$}\bar\l\bar\p^2$.
This formula is rather unwieldy, however, as we are only interested
in the leading logarithms, it can be simplified somewhat.
Consider the $\hbar ln \bar H/\bar G$ terms in eq. (3.11). We can write
$$\hbar\ln{\bar H\over{\bar G}}=\hbar\ln{H\over G}
+\hbar\ln{\bar H\over{H}}-\hbar\ln{\bar G\over G}.\eqno(3.12)$$
Now the point is that the second and third terms on the right hand
side of eq. (3.12) do not contribute to the leading logarithms, since if
$\hbar\ln[(\bar{m^2}+\h\bar\l\bar\p^2)/H]$ is expanded in powers of $t$,
all the terms will be of the form $\hbar^{n+1}\ln^n(G/\m^2)$ which
are sub-leading logarithms. Thus, in the leading logarithmic
approximation we are entitled to make the replacement
$\ln(\bar H/\bar G)\rightarrow \ln(H/G)$ in eq. (3.11), ie.
$$\eqalign{
\left(1-{3\bar\l(t)\hbar\over{2(4\pi)^2}}\ln
{\bar H\over{\bar G}}\right)\rightarrow
&\left(1-{3\bar\l(t)\hbar\over{2(4\pi)^2}}\ln{H\over G}\right)\cr
&=\left(1-{(N-1)\l t\over{3(4\pi)^2}}-{3\l\hbar \over{2(4\pi)^2}}
\ln {H\over{\m^2}}\right)\left(
1-{(N+8)\l t\over{3(4\pi)^2}}\right)^{-1}.\cr}
\eqno(3.13)$$
Inserting eqs. (3.9) and (3.13) into eq. (3.11),
 the leading logarithms in $O(N)$ $\p^4$ theory sum to

$$\eqalign{
L=&{\l\over{24}}\p^4\bbr^{-1}\cr
&+{1\over2}m^2\p^2\bbr^{-1/3}
\br^{-{2\over3}{N-1\over{N+8}}}\cr
&-{m^4\over{2\l}}\left[
\bbr^{1/3}\br^{-{4\over3}{N-1\over{N+8}}}\right.\cr
&\left. -4{N-1\over{N-4}}\br^{-
{N-4\over{N+8}}}+{3N\over{N-4}}\right]+\L,\cr}\eqno(3.14)$$
where
$$s={\hbar\over2}\ln{H\over{\m^2}}\qquad\hbox{and}\qquad
t={\hbar\over2}\ln{G\over{\m^2}}.$$
Note that this result must be considered as a conjecture,
 since its derivation relied on eq. (3.7) which is
not proven here. Although we are unable to prove eq. (3.14), the reader can
easily verify that it is correct at the tree, one-loop and two-loop
level.

   If $m^2<0$ then the \sl tree \rm level minimum of the EP is given
by $\p^2=-6m^2/\l$ ie. at $G=0$. It is clear from eq. (3.6), that
some of the leading logarithms are not well behaved in the limit
$G\downarrow 0$. In particular the two-loop contribution
proportional to
$$-H\ln^2{G\over{\m^2}}+2H\ln{H\over{\m^2}}\ln{G\over{\m^2}},$$
diverges as $G\downarrow 0$. In fact, these divergences in $V^{(2)}$
are cancelled by infrared divergences in the non-logarithmic part
of $V^{(2)}$ [11]. Of course, just because individual terms
in $L$ are divergent in this limit, this does not imply that
$L$ as a whole is also divergent in this limit. If we let
$t\rightarrow -\infty$ in eq. (3.14), then the first two
terms vanish, however the behaviour of the remaining terms is
more complicated. If $N\leq 4$ then $L$ diverges, but
if $N>4$ we are left with a peculiar finite term
$$L(t\rightarrow-\infty)=-{3Nm^4\over{2(N-4)\l}}+\L,
\qquad N>4.\eqno(3.15)$$
Note that this is independent of $\hbar$, yet is \sl not \rm
(except in the limit $N\rightarrow \infty$) equal to the classical vacuum
 energy density,
which is given by
$$\rho_{\hbox{classical}}= -{3m^4\over{2\l}}+\L,\eqno(3.16)$$
in the case $m^2<0$.

An alternative approximation to the loop or leading logarithms
expansion is the large $N$ expansion [10]. The first order term in this
approximation amounts to  summing up the leading terms in $N$ at each
order in  perturbation theory.
$$\eqalign{V_1=&{\l\over{24}}\p^4+{1\over2}m^2\p^2+{\hbar NG^2\over{4(4\pi)^2}}
\left(\ln{G\over{\m^2}}-{3\over2}\right)+{\hbar^2 N^2 \l G^2\over{24 (4\pi)^4}}
\left(\ln{G\over{\m^2}}-1\right)^2\cr
&+\hbox{higher order terms,}\cr}\eqno(3.17)$$
where the term proportional to $N$ is just the leading contribution
to eq. (3.3), and the $N^2$ term is due to the two-loop \lq\lq figure of
eight'' graph.
Note that  only bubble graphs  with no $H$ propagators contribute to
$V_1$.
There is an exact (although implicit) expression for $V_1$
[13], which in \M reads
$${\partial V_1\over{\partial \p}}=\chi \p,\eqno(3.18)$$
where $\chi$ satisfies the gap equation
$$\chi=G+{\hbar\l N\chi\over{6(4\pi)^2}}\left(
\ln{\chi\over{\m^2}}-1\right).\eqno(3.19)$$
In fact, it is possible to derive the above expression using the RG;
one simply notes that all contributions beyond the tree level
to $ V'_1=\pd V_1/\pd \p$ will be proportional to the one-loop tadpole
with a $G$ propagator, or $V'_1-V'_{\hbox{tree}}\propto
G(\ln(G/\m^2)-1)$, so if we set $\m^2=G/e$ all loop terms in $V'_1$ vanish!
 ie. $V'_1(\m^2=G/e)=V'_{\hbox{tree}}=
\p G$. If we solve the RG for $V'$\footnote{$\dagger$}
{In general the RG for $V'$ is ${\cal D}V'=\gamma V'$ and so $V'=\bar
V'$ only if $\gamma=0$.}; $V'(\l,m^2,\p,\m)=V'(\bar \l(t),
\bar m^2(t),\p,\bar\m(t))$, now choose $t$ such that $\bar\m^2(t)=\bar
G(t)/e$ (this choice is just the \lq\lq gap'' equation (3.19)) so
that $V'=\bar V'(t)=\p \bar G(t)$, and since $V_1$ is made up of products of
one-loop graphs we only need use the one-loop RG functions,
ie. $\b_\l=\hbox{${1\over3}$}\hbar N\l^2/(4\pi)^2$,
$\b_{m^2}=\hbox{${1\over3}$}\hbar N \l m^2/(4\pi)^2$, $\gamma=0$
where non-leading terms in $N$ have been dropped. If we take
the large $N$ \sl and \rm the leading logarithmic approximation
(which amounts to making the less implicit choice
$t=\h\hbar\ln(G/\m^2)$ when solving the RG) one finds that
$$L_1=\left[{1\over2}m^2\p^2+{\l\over{24}}\p^4\right]
\left(1-{N\l t\over{3(4\pi)^2}}\right)^{-1}+\Lambda,\eqno(3.20)$$
which agrees with eq. (3.14) in the large $N$ limit.
The effective potential has  been computed to the next order in
the large $N$ expansion [14],  however the formula is not
very managable (although it may be possible to extract just the
leading logarithms and compare with eq. (3.14)).

To summarize, we have an expression for the leading logarithms sum
with the following properties:
\item{i)} For $N=1$, it reduces to the known result.
\item{ii)} In the large $N$ limit, it reduces to the known result.
\item{iii)} It is correct through to two-loops.
\item{iv)} It has the correct $\ln\m^2$ dependence.

Property i) is a consequence of the proposed boundary conditions
eq.(3.7),
while property iv) is a just a statement that the improved potential
was constructed using the one-loop running parameters ie. eqs. (3.9).
The reader might enquire whether it is possible to write down an
alternative improved potential with the above properties. The answer
is yes, but the result will look unnatural. For example,
 the first entry in eq. (3.14) (the $\p^4$ term) could be
replaced in the following way
$${\l\p^4\over{24}}\lbbr^{-1}\rightarrow
{\l\p^4\over{24}}\lbbr^{-1}
\left(1-\hbox{${(N-1)\l^3(s-t)^3\over{(4\pi)^6}}$}\right)^p,$$
without affecting the above properties (here $p$ is just a constant).

\line{\hfill}
\line{\hfill}

\centerline{\fourteenpoint\bf 4. $O(N)$ Yukawa theory}
\line{\hfill}
Here we repeat the calculation of the previous section for the
$O(N)$ Yukawa model. This is defined by the Lagrangian
$${\cal
L}={1\over2}(\d_\m\p)^2-{1\over2}m^2\p^2-{\l\over{24}}\p^4-\L+
\bar\psi_i( i\d\!\!\!/ -g\p)\psi_i,\eqno(4.1)$$
where $\psi_i\quad(i=1,..,N)$ is a $N$-component Dirac field.
We have $N$ massless (Dirac) fermions interacting with a scalar
field $\p$ via an $O(N)$ invariant
 Yukawa coupling (here $\p$ is an $O(N)$ singlet).
The tree level potential is
$$V^{(0)}={1\over2}m^2\p^2+{\l\over{24}}\p^4,\eqno(4.2)$$
and the one-loop potential is given by
$$V^{(1)}(\p)={\hbar H^2\over{4(4\pi)^2}}\left(
\ln{H\over{\m^2}}-{3\over2}\right)-{N\hbar F^2\over{(4\pi)^2}}\left(
\ln{F\over{\m^2}}-{3\over2}\right),\eqno(4.3)$$
where
$$H=m^2+\h\l\p^2,\qquad F=g^2\p^2.\eqno(4.4)$$
Once again we have two distinct logarithms to sum. The one loop RG
functions are
$$\eqalign{
\k\b^{(1)}_\l=&3\l^2+8N\l g^2-48Ng^4,\cr
\k\b^{(1)}_{m^2}=&(\l+4Ng^2)m^2,\cr \k \g^{(1)}=&2Ng^2,\cr}\qquad
\eqalign{
\k\b^{(1)}_g=&(2N+3)g^3,\cr
\k\b_{\L}^{(1)}=&\h m^4,\cr&\cr }\eqno(4.5)$$
where $\k=\f/\hbar$.
The one-loop running parameters are (see for example [15])
$$\eqalign{
\bar\p(t)&=\p B(t)^{2N\over{4N+6}},\quad
\bar g^2(t)=g^2/B(t),\cr
\bar\l(t)&=g^2{
a(\l-bg^2)B(t)^{{3a\over{4N+6}}-1}-b(\l-ag^2)B(t)^{{3b\over{4N+6}}-1}\over
(\l-bg^2)B(t)^{{3b\over{4N+6}}}-(\l-ag^2)B(t)^{{3b\over{4N+6}}}},\cr
\bar m^2(t)&=m^2B(t)^{-{4N\over{4N+6}}}\left[
{(\l-bg^2)B(t)^{3a\over{4N+6}}-(\l-ag^2)B(t)^{3b\over{4N+6}}\over
(a-b)g^2}
\right]^{-{1\over{4N+6}}},\cr
\bar\L(t)&=\L+{1\over{2\f}}\int^t_0dt'\bar m^2(t'),\cr}\eqno(4.6)$$
where
$$B(t)=\left(1-{(4N+6)g^2t\over\f}\right),\eqno(4.7)$$
and $a$ and $b$ are the roots of the quadratic equation
$$3y^2+(4N-6)y-48N=0.\eqno (4.8)$$
In order to sum the leading logarithms in this theory we solve the
RG with suitable boundary conditions at $\m^2=F$. We \sl assume \rm that at
$\m^2=F$ the leading logarithms reduce to the $N=0$ form (eq.
(2.12)). That is we take
$$\eqalign{L(\m^2=F)=&
{\l\over{24}}\p^4\left(1-{3\l\hbar\over{2\f}}\ln{H\over
F}\right)^{-1}+
{1\over2}m^2\p^2\left(1-{3\l\hbar\over{2\f}}\ln{H\over
F}\right)^{-1/3}\cr
&-{m^4\over{2\l}}\left(1-{3\l\hbar\over{2\f}}\ln{H\over F}\right)^{1/3}+
{m^4\over{2\l}}+\L.\cr}.\eqno (4.9)$$
Choosing
$$t={\hbar\over2}\ln{F\over\m^2},\eqno(4.10)$$
and proceeding in the same way as the $O(N)$ scalar calculation in
the previous section, the leading logarithms sum to
$$\eqalign{L=&{\bar\l(t)\over{24}}\bar\p^4(t)\left(
1-{3(s-t)\bar\l(t)\over\f}\right)^{-1}
+{1\over2}\bar
m^2(t)\bar\p^2(t)\left(1-{3(s-t)\bar\l(t)\over\f}\right)^{-1/3}\cr
&-{\bar m^4(t)\over{2\bar\l(t)}}\left(
1-{3(s-t)\bar\l(t)\over\f}\right)^{1/3}+
{\bar m^4(t)\over{2\bar\l(t)}}
+{1\over2\f}\int^t_0dt'\bar m^4(t')+\L,\cr}\eqno(4.11)$$
where
$$s={\hbar\over2}\ln{H\over{\m^2}}\quad\hbox{and}\quad
t={\hbar\over2}\ln{F\over{\m^2}}.$$
As in the $O(N)$ scalar case we discarded non-leading
logarithms. At this point it is appropriate to compare this
calculation with the work of ref. [7].
In their treatment of the Yukawa model they considered two cases:

\item{i)} $m^2<<F$.
\item{ii)} $m^2>>F$.

In case i) they noted that since $\ln(H/F)$ can be considered small,
one is entitled to write $\ln(H/\m^2)=\ln(F/\m^2)+\ln(H/F)$, and sum
up the $\ln(F/\m^2)$ terms in the usual way (ie. using the method
reviewed in section 2).

In case ii) their treatment had similarities to the calculation
presented here. Their improved potential was obtained by solving the
RG with boundary conditions specified at $\m^2=F$. They also noted
that one could not use the tree potential as a boundary condition,
and as  here they employed ``improved'' boundary conditions.
Their boundary condition for the leading logarithms read
$$L(\m)={\tilde\l\over{24}}\tilde\p^4+{1\over2}\tilde m^2\tilde\p^2+
\tilde\L\quad\hbox{at}\quad\m^2=F,\eqno(4.12)$$
where
$$\eqalign{\tilde \l=&\l+{3\l^2\hbar\over{2\f}}\ln{m^2\over{\m^2}},
\quad \tilde m^2=m^2+{\l m^2\hbar\over{2\f}}\ln{m^2\over{\m^2}},
\quad \tilde\p=\p\cr
\tilde \L=&\L-{m^4\hbar\over{4\f}}\ln{m^2\over{\m^2}}.\cr}\eqno(4.13) $$
The choice of boundary condition was motivated by ideas from
effective field theory, where the scalar particle is treated
as a very heavy particle of mass $m$. The effects of the heavy
particle lead to a shift of the parameters of the low-energy
theory. However, it is arguable that eq. (4.12) is not really
valid for $\m^2=F$, since in the regime $m^2>>F$ terms of the form
$\hbar^2\ln^2(m^2/F)$, $\hbar^3\ln^3(m^2/F)$, etc. (which
are neglected in eq. (4.13)) will be large, and should
be included in any calculation of the leading logarithms.
Although the boundary condition used in ref. [7] seems to be
questionable for $H>>F$, it must be emphasized that the formula
presented here (eq. (4.11)) may also break down at some \sl higher \rm
order in perturbation theory. This is because our boundary condition
(eq. (4.9)) is a conjecture, which does hold at the tree, one-loop
and two-loop level. It is plausible (but not certain) that it
survives to all orders in perturbation theory.
\line{\hfill}
\line{\hfill}

\centerline{\fourteenpoint \bf 5. Discussion}
\line{\hfill}

We have presented a calculation of the leading logarithms of the
EP for two simple theories. These theories are perhaps the simplest
renormalisable theories where the perturbation series contains more
than one logarithm. However, our calculations were based on an
unproven conjectured property of the leading logarithms, which
means they could break down at some power of Planck's constant.
Moreover, we were unable to (even in principle) sum the sub-leading
logarithms, since we have no corresponding conjecture concerning
the sub-leading logarithms (in contrast to the single scale case,
where it is known how to sum the sub-leading logarithms, and has
been done explicitly for $\p^4$ theory [5]). But it could be
that there is a more systematic procedure that will reproduce
eqs. (3.14) and (4.11), without the need for such conjectures.
One possibility would be to introduce extra renormalisation scales
[16], which lead to several RG equations. In this approach it may
be possible to use one of the \lq\lq partial'' RG equations to
obtain improved boundary conditions for the conventional RG
equation (although the beta functions will depend on the ratios
of the scales, beyond one loop). Alternatively, it may be possible
to justify the boundary conditions used here by a careful analysis
of the properties of the Feynman diagrams contributing to the
EP.

Despite the rather unsystematic nature of our procedure it is quite
easy to extend it to  theories with more than two logarithms. For
example, consider $N$ scalars interacting with $M$ fermions via an
$O(M)$ invariant Yukawa coupling (assume that in the absence of
the Yukawa term we also have the usual $O(N)$ invariant action
for the scalars). Now the EP for this theory will have three
logarithms ($\ln(H/\m^2)$, $\ln(G/\m^2)$ and $\ln(F/\m^2)$).
As a boundary condition for the RG we could use eq. (3.14)
at $\m^2=F$.
However things may not be so simple in gauge theories, since one
does not have total freedom to vary the number of massive gauge bosons and
scalars. This is because we must have sufficient Goldstone bosons
to \lq\lq feed'' the vectors with masses.

In the case of more realistic theories (such as the Standard
Model) one is forced to use
numerical methods, since the running parameter equations
 can be very
complicated (even at one-loop). The standard model effective
potential has several logarithms (five, if one neglects all
Yukawa couplings, except that involving the top quark), but in this
case it is not always necessary to sum them all separately. In refs.
[1,2,3] where bounds for the masses of the top and Higgs were
obtained via the assumption of electroweak vacuum stability, the
form of the EP was required in the region $\p>>M_Z$. In this regime
one certainly has large logarithms to sum, fortunately for these
large $\p$ values, the differences between the five logarithms
can be considered small (ie. all the logarithms are well approximated
by $t=\hbar\ln(\p/\m)$). Thus, in this case there is only one
logarithm to sum up, which can be done in the usual way. However,
many extensions of the standard model  do possess additional scales.
In such cases one could apply the methods described here by solving
the RG in the usual way \sl but with improved boundary conditions.
\rm These improved boundary conditions (analogous to eqs. (3.7)
and (4.9))  could be determined numerically.
\line{\hfill}
\line{\hfill}
\centerline{\fourteenpoint \bf Acknowledgements}
\line{\hfill}
\line{\hfill}

I am grateful to  B.Dolan,
J.Gracey, D.R.T.Jones, F.Krahe, L.O'Raifeartaigh and
C.Stephens
 for useful discussions. Thanks also to the referee for a
useful suggestion.
 \line{\hfill}
\line{\hfill}

\centerline{\fourteenpoint \bf References}
\line{\hfill}
\settabs 20\columns
\+\hfill[1]&& M.J.Duncan, R.Phillippe and M.Sher, Phys Lett. B153 (1985)
165;\cr
\+&& M.Sher and H.W. Zaglauer, Phys Lett. B206 (1988) 527; \cr
\+&&M.Lindner, M.Sher and Zaglauer, Phys Lett. B228 (1989) 139;\cr
\+&&M.Sher, Phys Rep. 179 (1989) 274;\cr
\+&&J.Ellis, A.Linde and M.Sher, Phys. Lett. B252 (1990) 203.\cr
\+\hfill[2]&& C.Ford, D.R.T.Jones, P.W.Stephenson and
M.B.Einhorn, Nucl. Phys.\cr
\+&& B395 (1993) 17.\cr
\+\hfill[3]&& M.Sher, Phys. Lett. B317 (1993) 159.\cr
\+\hfill[4]&& S.Coleman and E.Weinberg, Phys Rev. D7 (1973) 1888.\cr
\+\hfill[5]&& B.Kastening, Phys. Lett. B283 (1992) 287.\cr
\+\hfill[6]&& M.Bando, T.Kugo, N.Maekawa and H.Nakano, Phys. Lett.\cr
\+&&  B301 (1993) 83.\cr
\+\hfill[7]&& M.Bando, T.Kugo, N.Maekawa and H.Nakano, Prog. Theor.\cr
\+&&  Phys. 90,
(1993) 405.\cr
\+\hfill[8]&& H.Nakano and Y.Yoshida, Phys. Rev. D49 (1994) 5393.\cr
\+\hfill[9]&& B.W.Lee and J.Zinn-Justin, Phys. Rev. D5 (1972)
3121, Appendix B;\cr
\+\hfill[10]&&R.Jackiw, Phys. Rev. D9 (1974) 1686.\cr
\+\hfill[11]&& C.Ford and D.R.T.Jones, Phys. Lett. B274 (1992)
409,\cr
\+&&(erratum B285 (1992) 399).\cr
\+\hfill[12]&& B.Kastening, Renormalization group improvement of the\cr
\+&&effective potential in massive $O(N)$ symmetric $\p^4$ theory,\cr
\+&& UCLA preprint UCLA/92/TEP/26 (1992) (hep-ph 9207252).\cr
\+\hfill[13]&& H.J.Schnitzer, Phys. Rev. D10 (1974) 1800 and
2042.\cr
\+\hfill[14]&&R.G.Root, Phys. Rev. D10 (1974) 3322.\cr
\+\hfill[15]&& E.Elizalde and S.D.Odintsov, Renormalization-group
improved \cr
\+&&
effective potential for interacting theories with several mass\cr
\+&&
scales in curved spacetime,
 Barcelona preprint UB-ECM-PF 93/22 \cr
\+&&(hep-th 9401057).\cr
\+\hfill[16]&&M.B.Einhorn and D.R.T.Jones, Nucl. Phys. B230 (FS10)
(1984) 261.\cr
\vfill\end